# Optimized Embedded Implementation of Hyperspectral–Multispectral Image Fusion on Raspberry Pi

Salah Eddine Brezini
*Intelligent Systems Research Laboratory (Laresi)*
*University of Science and Technology of Oran*
Oran,Algeria
salaheddine.brezini@univ-usto.dz

Okba Bekhelifi
*Intelligent Systems Research Laboratory (Laresi)*
*University of Science and Technology of Oran*
Oran,Algeria
okba.bekhelifi@univ-usto.dz

Oussama Mezouar
*Intelligent Systems Research Laboratory (Laresi)*
*University of Science and Technology of Oran*
Oran,Algeria
oussama.mezouar@ensttic.dz

Chams Eddine Choucha
Advanced Data Science and Cognitive Applications Laboratory
*University of Science and Technology of Oran*
Oran,Algeria
chamseddine.choucha@univ-usto.dz

Sarra Boukhacheba
*Intelligent Systems Research Laboratory (Laresi)*
*University of Science and Technology of Oran*
Oran,Algeria

Fethi Abdelatif Dali
*Intelligent Systems Research Laboratory (Laresi)*
*University of Science and Technology of Oran*
Oran,Algeria

***Abstract*— Remote sensing optical images have become central to a wide range of applications. In particular, hyperspectral images, with their high spectral resolution, enable the extraction of rich information about the objects and materials present in the observed scene. Nevertheless, processing such data comes at the expense of a high computational load due to its large data volume, making real-time processing very difficult to achieve. Recently, we proposed an approach to investigate the feasibility of processing such data on a Raspberry Pi by implementing a hyperspectral super-resolution technique, namely HSB-SV. However, the implementation resulted in high computational time. To overcome this limitation, we apply computational optimization techniques based on migrating the most intensive operations to PyTorch and edge inference frameworks such as ONNX Runtime and ExecuTorch with the XNNPACK backend. The proposed optimized implementation is deployed on a Raspberry Pi 5 platform. Experimental results demonstrate a significant reduction in computational time, achieving a 1.48x overall speedup on the Raspberry Pi 5, the total execution time decreases significantly, from 527.1 ms to 356.7 ms corresponding to a 32.3% dminution, while fully preserving the fusion quality, making the approach more suitable for embedded and edge computing scenarios, particularly for UAV-based hyperspectral remote sensing applications.**

***Keywords—Hyperspectral imaging, multispectral imaging, image fusion, hyperspectral super-resolution, embedded systems, edge computing, Raspberry Pi 5, computational optimization, real-time processing.***

## I. Introduction

Raspberry Pi has emerged as a widely used low-cost, small-form-factor computing platform for a broad range of applications. It is no longer limited to simple data acquisition tasks, but also supports data processing and storage capabilities. In addition, it can be deployed as a lightweight computing platform for various embedded applications, including image processing [1] and precision agriculture [2]. Moreover, Raspberry Pi exhibits efficient energy consumption, with continuous development since its introduction in 2012, making it suitable for real-time embedded systems. Thus, Raspberry Pi is well-suited for platforms such as unmanned aerial vehicles (UAVs), since its lightweight design and low energy consumption are critical factors [3]. In this context, UAVs have become widely used platforms for remote sensing applications [4]. Remote sensing optical images are widely used in a variety of applications, including precision agriculture [5], environmental monitoring [6], [7] renewable energy [8] and urbanism [9], [10], due to their ability to provide valuable information about the Earth's surface [11]. In particular, hyperspectral images (HSIs) play a crucial role in remote sensing applications due to their high spectral resolution, which enables accurate material discrimination and detailed scene analysis [12]. However, this rich spectral information comes at the cost of a significant increase in data dimensionality, leading to substantial computational challenges for practical processing and real-time applications.

Recently, we implemented a lightweight hyperspectral super-resolution technique [13], namely Hyperspectral Super-Resolution with Spectral Bundles for Spectral Variability (HSB-SV) [14] , on a Raspberry Pi to validate the feasibility of processing hyperspectral images on such a low-cost platform. HSB-SV aims to fuse an HSI with a multispectral image (MSI) to overcome one of the main limitations of hyperspectral imaging, namely its low spatial resolution. This fusion process, also referred to as hypersharpening [15], [16], produces a high-spatial-resolution hyperspectral image by combining the rich spectral content of the HSI with the fine spatial detail of the MSI. The HSB-SV demonstrated its effectiveness and computational efficiency on both synthetic and real datasets [14], making it particularly well-suited for urban remote sensing scenarios where both high spectral resolution needed for accurate material discrimination and high spatial resolution required for precise mapping of complex urban structures are simultaneously required [10]. Despite the successful implementation on Raspberry Pi 3 and the encouraging results obtained, the processing time remained relatively significant for real-time and rapid applications. To address this limitation, this work proposes an optimized implementation of HSB-SV on Raspberry Pi 5. The Raspberry Pi 5 is equipped with a quad-core Arm Cortex A76 processor. At 2.4 GHz clock, it can deliver up to 19.2 double precision GFLOPS per core [24] thanks to the ARMv8.2 architecture and the Arm NEON Single Instruction, Multiple Data (SIMD) extension. Moreover, Deep Learning frameworks like PyTorch and inference engines provide highly accelerated execution against conventional mathematical libraries. The proposed optimization consists of migrating the computationally intensive operations of the original Python implementation to PyTorch and edge inference frameworks such as ONNX Runtime and ExecuTorch with the XNNPACK backend. By leveraging highly optimized CPU kernels, the proposed implementation reduces the execution time of HSB-SV while preserving the performance of the original method, thereby improving its suitability for embedded hyperspectral image processing applications.

The remainder of this paper is organized as follows. Section II describes the HSB-SV algorithm and its associated mathematical formulation. Section III presents the computational optimization of HSB-SV for embedded deployment. Section IV presents the experimental results. Finally, Section V concludes the paper.

## II. HSB-SV Algorithm

This section describes the mathematical modeling underlying the HSB-SV method and details the corresponding algorithm. HSB-SV extends the linear spectral unmixing (LSU) model to account for a physical phenomenon known as spectral variability (SV) [17], [18]. The LSU model represents each pixel of an HSI or MSI as a linear combination of pure spectral signatures, referred to as endmembers, weighted by their corresponding abundance coefficients, which represent the proportion of each material present within the pixel [19]. Thus, for the rest of this paper the considered HSIs and MSIs are described by:

$$X_h = A_h S_h, \tag{1}$$

$$X_m = A_m S_m, \tag{2}$$

where $X_h \in R_+^{P_h \times L_h}$ and $X_m \in R_+^{P_m \times L_m}$ are the observed hyperspectral and multispectral images, respectively. $P$ stands for the number of pixels and $L$ the number of spectral bands, with $h$ and $m$ indices referring to the hyperspectral and multispectral images, respectively. $A_h \in R_+^{P_h \times N}$ and $A_m \in R_+^{P_m \times N}$ are the spatially degraded associated abundance fractions of the HSI and associated abundance fractions of the MSI. The estimated hyperspectral and spectrally degraded multispectral endmember spectra are denoted by $S_h \in R_+^{N \times L_h}$ and $S_m \in R_+^{N \times L_m}$. The spatially degraded abundance $A_h$ and the spectrally degraded endmember spectra $S_m$ are here modelled as

$$A_h = F A_m, \tag{3}$$

$$S_m = S_h R. \tag{4}$$

the matrix $F \in R_+^{P_h \times P_m}$ represents the point spread function (PSF) and $R \in R_+^{L_h \times L_m}$ describes the spectral response function (SRF).

HSB-SV is built around two main stages. The first stage aims to estimate the spectral information of the HSI by constructing a spectral library, also referred to as spectral bundles, that captures the SV of the materials present in the scene. SV can be defined as the phenomenon whereby the spectral signature of a given material may vary depending on its spatial location within the considered HSI [18]. To account for SV, HSB-SV employs a simple strategy known as Automated Extraction of Endmember Bundles (AEEB). The AEEB technique consists of repeatedly applying an Endmember Extraction Algorithm (EEA) to randomly selected subsets of the HSI in order to build a spectral library that captures the variability of the materials present in the scene. The employed EEA method is the Vertex Component Analysis (VCA) [20]. The spectra bundles can be expressed as [21]

$$B = [B_1 | B_2 | \cdots | B_j] \tag{5}$$

where $B_j \in R_+^{L_h \times Y_j}$ denotes bundle representing the $j-$th class, $J$ represents the number of classes, $Y_j$ is the number of pure spectra in the $j-$th class and $Y$ the total number of endmember spectra of all classes with $Y = \sum_{j=1}^{J} Y_j$. Hence, an observed hyperspectral pixel spectrum $x_{h_i} \in R_+^{L_h \times 1}$ is expressed as

$$x_{h_i} = B a_{h_i} \tag{6}$$

where $a_{h_i} \in R_+^{Y \times 1}$ stands for the abundance coefficients corresponding to each individual spectrum of the endmember bundles $B$.

In the second stage, the spatial information is extracted from the MSI through a sparse regression technique namely sparse unmixing by variable splitting and augmented lagrangian (SUnSAL) based on the previously estimated spectral library and the considered MSI. First, The resulting spectral library from the first step is spectrally degraded using (4) to obtain $B_m$in order to match the spectral characteristics of the MSI sensors. As matter of fact the SRF is estimated during the fusion process by considering the method described in [22] , allowing HSB-SV to operate in a fully blind manner. SUnSAL is applied between the spectrally degraded $B_m$ and $X_m$ to obtain the high-spatial-resolution abundance fractions. The main objective of sparse unmixing is to estimate the abundance coefficients from a large spectral dictionary. Thus, the objective of SUnSAL is to estimate the high-spatial-resolution abundance fractions $a_{m_i}$ by optimizing the following cost function, separately for each pixel as

$$\min_{a_{m_i}} \frac{1}{2} \left\| B_m a_{m_i} - x_{m_i} \right\|_2^2 + \lambda \left\| a_{m_i} \right\|_1 \tag{7}$$

where $\|\cdot\|_2^2$ is the $\ell_2$-norm and $\|\cdot\|_1$ is the $\ell_1$-norm which is responsible for promoting sparsity. $\lambda$ is a non-negative parameter which tunes the relative weight between $\ell_1$ and $\ell_2$. SUnSAL utilizes the Alternating Direction Method of Multipliers (ADMM) to optimize (7). To help the reader to understand the SUnSAL method, they may refer to [23]. Given that the proposed work emphasizes time efficiency, sparse-based methods are particularly suitable due to their relatively low computational cost. Therefore, the combination of AEEB and SUnSAL is well tailored for this work and time efficiency purpose.

The last part of the proposed framework focuses on reconstructing the high-spatial-resolution HSI by integrating the spectral information extracted from the HSI with the spatial details provided by the MSI. Thus, the last stage of the proposed approach consists in creating the fusion product $\widetilde{X_f}$ by recombining the obtained matrices. Hence, each pixel spectrum $\widetilde{x_{f_i}}$ of the unobservable sharpened high-spatial-resolution hyperspectral image $\widetilde{X_f}$ is defined as

$$\widetilde{x_{f_i}} = Ba_{m_i}. \quad (8)$$

The complete diagram, composed of the three main parts of the HSB-SV, is illustrated in Fig. 1.

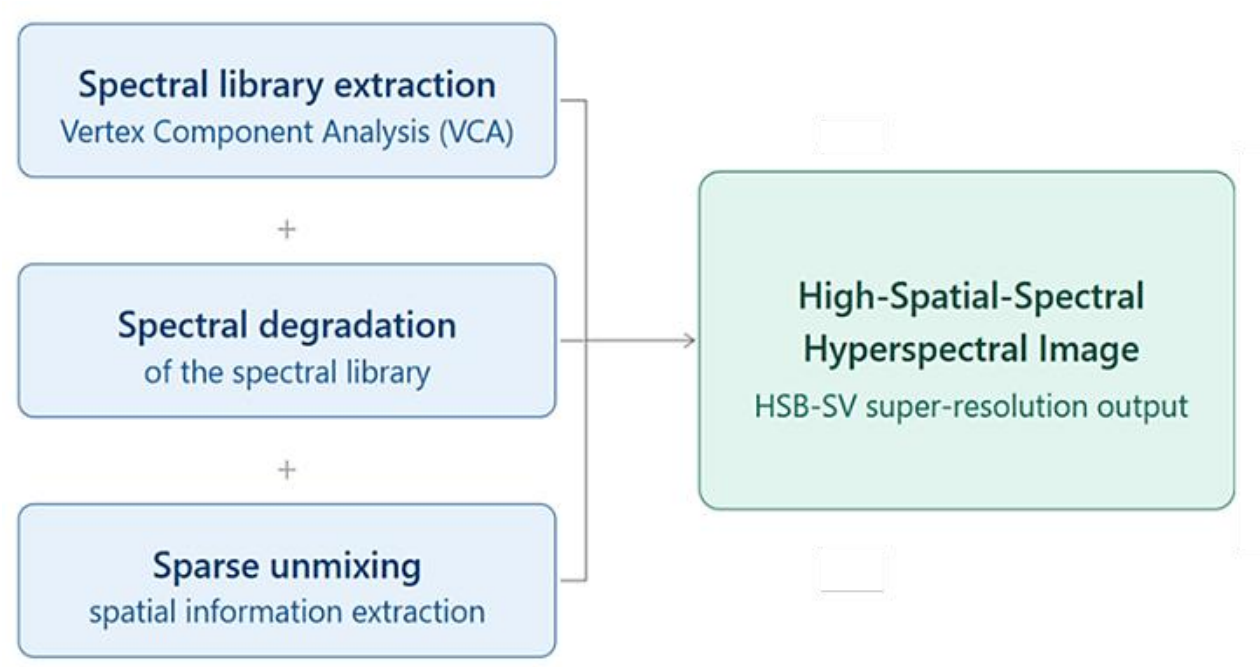

**Fig. 1.** Block diagram of the proposed HSB-SV

## III. Computational Optimization of HSB-SV for Embedded Deployment

### A. Accelerators

PyTorch ecosystem: One factor in the success of deep learning is the continuous development of open source frameworks that provide seamless training and deployment on different devices from dedicated GPUs to consumer computers and edge CPUs. PyTorch represents the computations as a dynamic computational graph at runtime, capturing the operations as a directed acyclic graph (DAG) where nodes denote tensor operations and edges track data dependencies [25]. When executed on CPUs, PyTorch invoke hardware specific optimized micro-kernels. For general linear algebra routines, these kernels use CUDA primitives on NVIDIA GPUs, whereas on CPUs, they utilize Intel onneDNN for x86-64 systems and OpenBLAS for ARM devices. Once the training is completed, the dynamic graph can be compiled into an optimized static graph in an intermediate representation (IR) format using the ahead-of-time (AOT) export feature. In this manner, the python interpreter overhead is bypassed [26].

To deploy such models on resource-constrained edge devices, the PyTorch ecosystem provides ExecuTorch (ET) as a native edge-deployment solution. ExecuTorch [27] lowers the IR into hardware agnostic edge dialect, and to use the hardware specific extensions, it partitions and delegates subgraphs to the hardware backend. For Arm CPUs, XNNPACK is the designated backend. XNNPACK [28] is an open-source, highly optimized library of floating-point neural network inference operators. The library achieves state-of-the-art efficiency on CPUs by mapping the localized computation graph directly onto parallel hardware execution pipelines. The highest improvements with these tools are observed for weight-quantized models and for mini-batches tensors inference as in Deep Learning use cases, however, we harness the optimized primitives in a stateless manner i.e. forward computation graphs with function calls without persistent trainable parameters.ONNXruntime (ORT): The Open Neural Network Exchange (ONNX) ecosystem enforces a standardized, model representation that encapsulates deep learning architectures into a serializable, framework-agnostic computational graph. As a high-performance inference engine built to execute this format, ONNX Runtime ingests this static directed acyclic graph (DAG) to apply graph optimization, such as constant folding, algebraic simplifications, and structural operator fusion [29]. To efficiently execute these optimized graphs, ONNX Runtime abstracts hardware-specific acceleration through Execution Providers (EPs) [30]. The engine evaluates the capabilities of the platform and partitions the global computation graph into hardware-compatible subgraphs [31]. Subgraphs assigned to the native CPU Execution Provider are then lowered into high-performance architecture primitives. In this way, ONNXRuntime ensures portable and stable low-latency execution.

### B.Optimization approach

Based on the analysis of the original Python implementation, the computational cost is dominated by intensive linear algebra operations, including matrix multiplications, pseudo-inverse computations, singular value decomposition (SVD), and non-negative least-squares (NNLS) optimization. Accordingly, an optimized implementation of HSB-SV is proposed. The optimization consists of first, porting the most computationally demanding operations to PyTorch and exporting the modules to the accelerators format. A mixed-design strategy was adopted, combining framework-supported operations with existing implementations for functions that are not natively supported by current inference engines. In particular, SVD and pseudo-inverse computations were migrated to PyTorch only, while the SciPy implementation of NNLS was retained as it uses the Lawson-Hanson active-set algorithm [32].

Since most of the computational burden is concentrated in the VCA and SUnSAL algorithms, these components were selected as the primary optimization targets. In addition to the framework-level optimizations, a minor yet effective algorithmic modification was introduced in the VCA component. In its original formulation [20], VCA estimates the SNR from the input data in order to select the appropriate projection strategy. However, in the context of UAV-based acquisition which is the target deployment scenario of the Raspberry Pi 5 platform considered in this work the acquisition distance is significantly shorter than that of satellite sensors, resulting in a higher and more stable SNR. Consequently, the SNR estimation step was bypassed by fixing the SNR to a constant value of 30 dB. This assumption eliminates the noise estimation overhead and reduces the computational cost of the VCA component without any meaningful loss of accuracy for the considered deployment scenario.

Six implementation configurations were evaluated: the original NumPy-based implementation, a full PyTorch migration, ONNX Runtime, ExecuTorch with the XNNPACK backend, and two hybrid configurations combining ONNX Runtime or ExecuTorch with PyTorch for operations not supported by inference engines. We run two experiments: first, we tested the conversion of VCA and SUnSAL in terms of error between the original code and new implementations and the execution time of each function separately. Second, we tested the entire HSB-SV method with the six configurations. In the first experiments, the results are reported as a mean of 50 iterations, and in the second experiments, we run 5 warm-up iterations to eliminate the influence of the initialization lag, then we run the HSB-SV process for 50 iterations and report the mean and standard deviation of these repetitions. All the tests were evaluated on two machines, a laptop (PC) with a 16 GB RAM and an Intel CPU i5-8365U @ 1.60 GHz Max Turbo 4.10 GHz having SIMD extensions: SSE, SSE2, SSE3, SSES3, SSE4_1, SSE4_2, FMA, AVX and AVX2. The target machine was a 16 GB RAM Raspberry Pi 5 equipped with an Arm Cortex-A76 @2.4 Hz having the SIMD extensions: ASIMD, FPHP, ASIMDHP and ASIMDRDM.

## IV. Experimental Results and Performance Evaluation

A benchmark synthetic dataset was employed to evaluate the proposed approach. Following Wald's protocol, the dataset was generated from a reference hyperspectral image (REF) [11] (Fig. 2), consisting of 100 × 100 pixels and 144 spectral bands covering wavelengths from 0.38 to 1.05 μm. The scene contains seven distinct materials and exhibits SV, making it suitable for assessing fusion algorithms under realistic conditions. To create the low-resolution hyperspectral image (LR-HSI), the reference image was spatially degraded using a Gaussian filtering operation and a downsampling factor of 4 [17]. The corresponding high-resolution multispectral image (HR-MSI) was synthesized using the spectral response functions of the WorldView-2 sensor, yielding 8 multispectral bands. This dataset was selected because it provides a well-established evaluation framework and has been extensively used to validate the effectiveness of the HSB-SV method on both synthetic and real hyperspectral data.

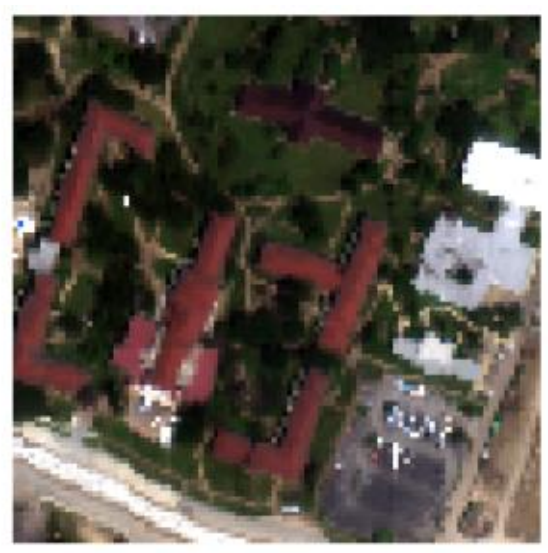

**Fig. 2** RGB representation of the reference image used in the conducted experiments

Table 1. shows the results of the first experiment for VCA. On PC, the MAE between the original code (Numpy) and the conversions is on the order of 1e-6 and the difference between PyTorch and the rest of accelerators is in the order of 1e-8. The former difference is due to the half precision (float32) operations of PyTorch, and the latter minimal difference is due to the shared backend primitives executed and they differ in the graph layout, memory planning and input data types. ExecuTorch inputs are torch tensors, whereas ONNXRuntime is fed with Numpy ndarrays. The same patterns is observed on the Pi 5, with a negligible increase. On the execution time, native PyTorch outperforms all the rest with a 14.82 millisecond (ms) (2.52x faster than the original), this is due to the use of the oneDNN library. ExecuTorch + XNNPACK comes second and ONNXRuntime comes third. On the Pi 5, ExecuTorch+XNNPACK is the fastest with 24.05 ms (1.4x improvement). Here, the absence of the AVX2 and FMA is seen. Nevertheless, on VCA, the three accelerators are competitive on both machines.

TABLE I. The results of the VCA conversion. For each row the Lib1-Lib2 represents the MAE difference between them.

| VCA | PC | Pi 5 |
|---|---|---|
| MAE (lower is better) | | |
| Original-PyTorch | 1.5189 e-06 | 2.6668 e-06 |
| Original-ORT | 1.5187 e-06 | 2.6667 e-06 |
| Original-ET | 1.5185 e-06 | 2.6669 e-06 |
| PyTorch-ORT | 1.5545 e-08 | 1.5972 e-06 |
| PyTorch-ET | **1.2600 e-08** | **1.1590 e-06** |
| RT-OT | 2.0300 e-08 | 1.6780 e-06 |
| Execution time (ms) (lower is better) | | |
| Original | 37.4225 | 33.6947 |
| PyTorch | **14.8186** | 26.2564 |
| ORT | 15.6846 | 25.6340 |
| ET | 15.1057 | **24.0489** |

Table 2. reports the results of the first experiment for SUnSAL. Here, the MAE is smaller on both machines with an order of 1e-7 for the difference between the original code and the conversions, and between the accelerators, the same minimal scale of 1e-8 is observed. In the execution time, the range is higher, with PyTorch showing the best performance of 52. 72 ms on PC and 170.53 ms on Pi 5, making an improvement of 6.80x and 1.53x respectively. The other accelerators fails to deliver competitive results due to the iterative nature of the algorithm, that creates a significant boundary overhead by passing data from native tensors to backend engines repeatedly. This overhead is absent in VCA because its iterations are smaller depending on the number of endmembers.

TABLE II. The results of the SUnSAL conversion. For each row the Lib1-Lib2 represents the MAE difference between them.

| SUnSAL | PC | Pi 5 |
|---|---|---|
| MAE (lower is better) | | |
| Original-PyTorch | 2.6343 e-07 | 1.0176 e-07 |
| Original-ORT | 2.6902 e-07 | 1.1421 e-07 |
| Original-ET | 2.6327 e-07 | 1.0903 e-07 |
| PyTorch-ORT | 9.3201 e-08 | **4.1215 e-08** |
| PyTorch-ET | 9.3201 e-08 | **4.1215 e-08** |
| RT-OT | **9.1164 e-08** | 5.8749 e-08 |
| Execution time (ms) (lower is better) | | |
| Original | 358.7010 | 261.4005 |
| PyTorch | **52.7196** | **170.5305** |
| ORT | 144.4345 | 193.8608 |
| ET | 205.3248 | 254.1831 |

The final experiment results is presented in Table 3. Here again, a clear win for PyTorch on PC is seen in line with the results of the first experiments.

TABLE III. HSB-SV EXECUTION TIME ON THE MACHINES AND 6 CONFIGURATIONS. STD: STANDARD DEVIATION, TIME IN SECONDS.

| Configutation / Machine | PC | | Pi 5 | |
|---|---|---|---|---|
| | Mean | Std | Mean | Std |
| Original | 0.5656 | 0.1278 | 0.5271 | 0.0678 |
| PyTorch | **0.1030** | 0.0113 | 0.3774 | 0.0816 |
| ORT | 0.2320 | 0.0268 | 0.3949 | 0.0538 |
| ET | 0.3553 | 0.0866 | 0.4979 | 0.0664 |
| ORT + PyTorch | 0.1046 | **0.0087** | 0.3663 | **0.0479** |
| ET + PyTorch | 0.1072 | 0.0123 | **0.3567** | 0.0677 |

The HSB-SV execution was improved by 5.49x marking 103 ms. The mixed accelerators functions is competitive to native PyTorch and the configuration ONNXRuntime (VCA) + PyTorch (SUnSAL) was the most stable with the least standard deviation. On Pi 5, the mixed configurations surprisingly surpassed PyTorch. As the SUnSAL execution had significant difference. The configuration Executorch+XNNPCAK (VCA) + PyTorch (SUnSAL) showed the fastest execution with 357 ms (1.4x improvement). This configuration took the best from each accelerator in line with experiment 1. These results show a clear advantage of the proposed optimization approach.

## V. CONCLUSION

This paper presentes an optimized embedded implementation of the HSB-SV hyperspectral and multispectral image fusion method on the Raspberry Pi 5, targeting edge computing deployment scenarios. To address the computational limitations identified in the original NumPy-based implementation, several optimization strategies were proposed and evaluated, based on migrating the most computationally intensive operations to PyTorch and inference engines such as ONNX Runtime and ExecuTorch with the XNNPACK backend. Experimental results demonstrate significant performance gains on the Raspberry Pi 5. At the individual component level, the VCA algorithm achieved its best acceleration with ExecuTorch+XNNPACK, reducing the execution time from 33.69 ms to 24.05 ms (1.4x speedup). More significantly, the SUnSAL algorithm the most computationally demanding component of the pipeline was accelerated by 1.53x using PyTorch, dropping from 261.40 ms to 170.53 ms. At the full pipeline level, the hybrid configuration combining ExecuTorch + XNNPACK for VCA and PyTorch for SUnSAL proved to be the most effective strategy on the Raspberry Pi 5, reducing the total execution time from 527.1 ms to 356.7 ms, representing a 1.48x overall speedup. This configuration successfully exploits the strengths of each accelerator, leveraging XNNPACK's optimized ARM SIMD primitives for VCA and PyTorch's efficient linear algebra kernels for SUnSAL. In all cases, the numerical accuracy of the original method was fully preserved, with mean absolute errors on the order of 1e-6 or lower, confirming that the optimizations introduce no meaningful degradation in fusion quality. These findings confirm that the proposed optimization approach successfully bridges the gap between the computational demands of hyperspectral image processing and the constraints of low-cost embedded platforms such as the Raspberry Pi 5. The optimized HSB-SV implementation is thus well-suited for real-time remote sensing applications on UAV-based systems, paving the way for practical, energy-efficient hyperspectral data processing at the edge. Several promising directions can be explored to further improve the proposed approach. First, the evaluation on larger and more diverse real-world datasets would provide a more comprehensive assessment of the method's robustness and scalability under practical remote sensing conditions. Second, a significant computational bottleneck that remains to be addressed lies in the spectral response function (SRF) estimation step, which contributes considerably to the overall processing time. Reducing or approximating this estimation for instance by leveraging prior knowledge of the sensor characteristics or adopting a lightweight calibration procedure could lead to substantial additional speedups, bringing the method even closer to true real-time performance on embedded platforms such as the Raspberry Pi 5.